\documentclass[%
 aip,
 amsmath,amssymb,
 reprint,%
]{revtex4-1}

\usepackage{graphicx}
\usepackage{dcolumn}
\usepackage{bm}

\usepackage[utf8]{inputenc}
\usepackage[T1]{fontenc}
\usepackage{mathptmx}
\usepackage{etoolbox}
\usepackage{textcomp} 
\usepackage{siunitx}
\usepackage{mhchem}
\usepackage{makecell}
\usepackage{booktabs}

\DeclareSIUnit{\atpercent}{at.\%}
\DeclareSIUnit{\rpm}{rpm}
\DeclareSIUnit\torr{Torr}
\DeclareSIUnit\sccm{sccm}
\usepackage{subcaption}
\makeatletter
\def\@email#1#2{%
 \endgroup
 \patchcmd{\titleblock@produce}
  {\frontmatter@RRAPformat}
  {\frontmatter@RRAPformat{\produce@RRAP{*#1\href{mailto:#2}{#2}}}\frontmatter@RRAPformat}
  {}{}
}%
\makeatother
\begin{document}

\preprint{AIP/123-QED}

\title[]{Ru Alloying in Ni/Al Reactive Multilayers: Experimental Observations and Molecular Dynamics Simulations}
\author{Nensi Toncich}%
\affiliation{ 
Laboratory for Nanometallurgy, Department of Materials, ETH Zürich, Switzerland
}%
\author{Ankit Yadav}
 \affiliation{Institute of Physics of Materials, The Czech Academy of Sciences, Brno, Czech Republic.}
 \affiliation{Central European Institute of Technology, Brno University of Technology, Brno, Czech Republic}
\author{Jan Fikar}
 \affiliation{Institute of Physics of Materials, The Czech Academy of Sciences, Brno, Czech Republic.
}%
\email{nensi.toncich@mat.ethz.ch}
\author{Ralph Spolenak}
 \affiliation{
Laboratory for Nanometallurgy, Department of Materials, ETH Zürich, Switzerland
}%
\date{\today}

\begin{abstract}
Reactive multilayer thin films, a class of energetic materials, are increasingly recognized for their potential in joining applications, utilizing the chemical energy released as heat during exothermic reactions. These materials hold also promise for additional diverse technological applications, which require precise control over heat release rates and reaction propagation velocities. The microstructural properties of reactive multilayers play a critical role in determining their chemical reaction behavior. Among these, Ni/Al reactive multilayers have been extensively studied and used due to their favorable characteristics. In this study, we explore the incorporation of ruthenium (Ru) as a co-alloying element with nickel (Ni) in the Ni/Al system to investigate its impact on the materials properties, with a particular focus on reaction velocity and temperature. Ru enhances the reaction rates, but also causes a composition dependent phase transition in the as-deposited state from fcc to hcp. Additionally, molecular dynamics simulations are employed to examine the effects of Ru co-alloying with Ni, providing deeper insights into the underlying mechanisms. This work aims to advance the understanding of Ru's role in influencing the performance of Al/Ni-based reactive multilayers for advanced applications.
\end{abstract}

\maketitle

\section{\label{Intro}Introduction:\protect\\ }

Over the past decades, reactive multilayer (RM) films have garnered significant interest because of their ability to produce fast, highly exothermic reactions. Multilayers composed of alternating bilayers of two elements with a high enthalpy of mixing can be ignited through localized heating, triggering a self-sustaining reaction.~\cite{Floro1986} A prominent example of such materials, known as reactive multilayer nanofoils (RMNFs), is the Ni/Al multilayer system~\cite{Knepper_2009}. 
These reactions serve as effective heat sources for applications such as low-temperature soldering and bonding in microsystems packaging~\cite{Fujii1_2013, Fujii_2013}. When sufficient heat is locally generated, it triggers a self-propagating reaction that spreads through the adjacent material, creating a self-sustaining reaction front. This front advances through the multilayer structure at specific temperatures and velocities~\cite{Fujii1_2013,Fujii_2013,GRAPES2016105}. Recently, the integration of reactive multilayer systems into microelectronics and micromechanical systems (MEMS) as a bonding technique has gained considerable momentum, drawing increased attention from the scientific community~\cite{Braeuer_2014,Boettge_2010}.
The reaction characteristics of Ni/Al reactive multilayer systems (RMS) have been widely studied due to their rapid heat release during self-propagating reactions. Key parameters such as maximum temperature and propagation velocity can be controlled by adjusting the bilayer thickness or modifying the degree of intermixing within the system~\cite{Gavens2000,LIU2023118167,Adams2018,toncich2025}.

However, modifying these parameters requires intricate adjustments to the fabrication process. Consequently, numerous studies have focused on developing new methodologies to regulate the heat release rate during the reaction. A wide variety of binary systems has been investigated, including aluminides such as Al/Pt~\cite{Adams2018}, Al/Ni~\cite{Dyer1995}, Al/Ru~\cite{ABOULFADL2019344}, and Al/Ti~\cite{GACHON20051225}, as well as other combinations such as Ni/Ti~\cite{Adams2009,Fedor15102025} and Nb/Si~\cite{REISS1999217}. Studies have demonstrated that incorporating alloying elements into multilayers can either increase or decrease reaction propagation velocities~\cite{Danzi2019}. Ni/Al multilayers were alloyed by substituting nickel with elements such as copper and platinum, significantly altering reaction front temperature and velocity. This approach enables controlled or rapid heat release for various engineering applications without changing the multilayer architecture. For instance, Fritz \textit{et al.} successfully achieved slower propagation rates by creating low-density compacts of multilayer particles~\cite{FRITZ20111084,Fritz2015}. Woll \textit{et al.} identified Ru/Al multilayers as a novel reactive material that uniquely combines high energy density with ductile reaction products, ideal for reliable joining. They reported reaction velocities up to 10.9 m/s and peak temperatures near 2000  \textdegree C, along with the formation of a ductile B2-RuAl phase confirmed by experiments and simulations~\cite{Woll2016}.

Although such experimental advances in ternary systems are promising, atomistic insight into these complex multilayers remains limited. Molecular Dynamics (MD) simulations have become a widely used tool to study reactive multilayers~\cite{Baras2018}. Although limited to relatively small simulation sizes, MD offers precise control over system geometry, composition, and initial conditions, enabling detailed investigations of diffusion behavior, crystallization pathways, and front propagation dynamics. Most existing MD studies have focused on the well-established Ni-Al system~\cite{POLITANO201525,TURLO2016189,ROGACHEV2016158}, uncovering key insights into how microstructural features—such as grain boundaries, grain size, and vacancy concentrations—influence the reaction front~\cite{Witbeck2020,Witbeck2019,fabian2021}. Recent work has also explored geometric variations, including nanocrystals and premixed interlayers, shedding light on how pre-existing atomic-scale mixing impacts reaction kinetics.

Despite growing experimental interest in ternary multilayer systems, such as Al–Ti–Si~\cite{SEN201725} and Al–Ru–Ni/Pt/Hf~\cite{Pauly2018}, detailed atomistic studies on such chemically complex systems are still limited. In this work, we bridge this gap by presenting a combined experimental and computational study of a novel ternary system comprising aluminum and nickel co-alloyed with ruthenium (Al/Ni(Ru)). Through MD simulations and experimental validation, we investigate how Ru incorporation alters the reactivity, front propagation velocity, peak temperatures, and resulting microstructure. Our study provides new insight into reactivity tuning via alloying, offering a promising route for designing advanced reactive multilayers with tailored thermal and mechanical behavior.

\begin{figure*}[ht!]
    \centering
    \includegraphics{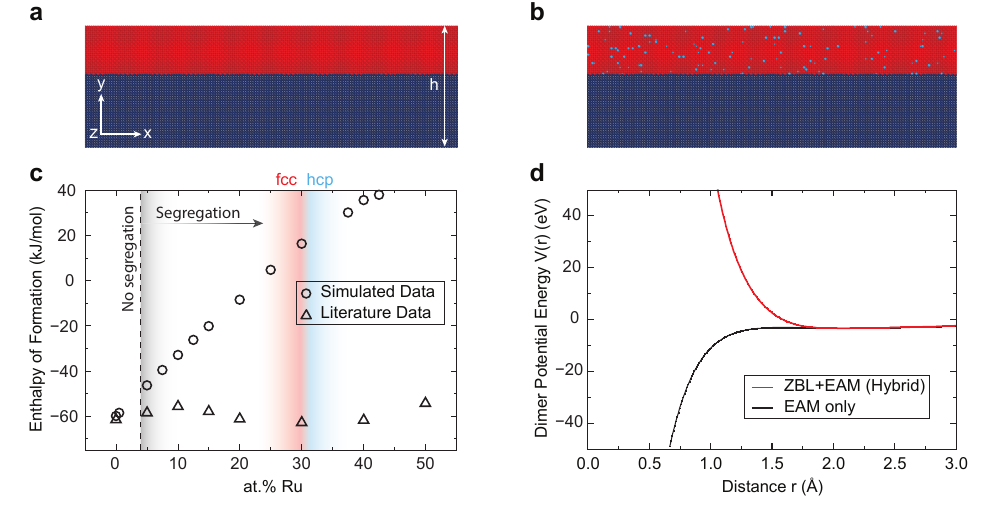}
    \caption{Initial multilayer samples with total thickness $h$=10\,nm, cropped along the $x$-direction. (a) Al/Ni sample containing 0 at.\% Ru. (b) Al/Ni sample with 3 at.\% Ru substitution in the Ni layers. Ni atoms are shown in red, Al atoms in blue, and Ru atoms in cyan. (c) Enthalpies of formation (EAM-derived) for Al(50 at.\%)/Ni–Ru compounds with varying Ru percentages in the overall Al/Ni-Ru composition, compared to literature values~\cite{SU2005217}. Error bars for the simulated values are smaller than the plotting symbols and are therefore omitted. Above the 4 at.\% Ru concentration we observe nonphysical Ru segregation. and absence of segregation. The shaded band indicates the fcc-to-hcp transition range identified from experimental XRD for the Ni–Ru layer in the as-deposited multilayers (Al remains fcc across the full composition range). (d) Comparison of the Ni–Ru dimer interaction using the EAM potential by Yue \textit{et al.}~\cite{YUE2020539} and the hybrid ZBL+EAM potential.}
    \label{fig:mdmlinitial}
\end{figure*}

\section{Experimental Section}

\subsection{Fabrication of Al/Ni-Ru Reactive multilayers}
Reactive multilayer films were synthesized using a commercial magnetron sputtering system (PVD Products Inc.). All depositions were performed in direct current (DC) mode under an argon atmosphere maintained at a pressure of \qty{3}{\milli\torr} with a flow rate of \qty{50}{\sccm}. Aluminum (99.99\%, MaTeck GmbH) was deposited at a constant power of \qty{250}{\watt}. The composition of the nickel-based layers was controlled through co-sputtering of Ni (99.99\%, MaTeck GmbH) and Ru (99.99\%, MaTeck GmbH) by adjusting the relative sputtering power. 

The multilayer architecture was designed with a bilayer thickness of \qty{50}{\nano\meter}, comprising a total of 50 bilayers. Each bilayer consisted of a \qty{30}{\nano\meter} Al layer and \qty{20}{\nano\meter} Ni-Ru layer. This strategy resulted in multilayers with an overall Al content of approximately 50–55 at.\%, while the Ni fraction decreased from ~50 at.\% to ~0 at.\% and the Ru fraction increased correspondingly from ~0 at.\% to ~45 at.\% across the sample series. The resulting targeted compositions of the deposited films are summarized in the supplementary material (STab.~1). The stacking sequence was maintained such that aluminum was always the first deposited layer, while the terminal layer was composed of Ni, Ni-Ru, or Ru, depending on the specific composition of the multilayer system. 

The base pressure prior to deposition was maintained below \qty{7e-7}{\torr} to ensure high purity of the films. During deposition, the substrate was rotated at a speed of \qty{30}{\rpm} to ensure film uniformity and its temperature was kept below \qty{40}{\degreeCelsius}.

The multilayer systems were deposited on two types of substrates: silicon (100) wafers coated with a photoresist (AZ 4533) and bare silicon wafers.

\subsection{Mechanical Characterization}
The mechanical properties of the as-deposited RM were measured on samples deposited directly on bare Si wafers using an iNano nanoindenter (Nanomechanics, Inc.) with an InForce50 actuator. All the samples were indented to a target load of 6 mN using a diamond Berkovich tip (SyntonMDP). Indents were created using the $\text{NanoBlitz}^{TM}$ program in a square grid of 10x10 indents at 3 locations per condition with a spacing of 3 µm between points. Elastic modulus values were extracted using the Oliver–Pharr method, assuming a Poisson’s ratio of 0.34 for the RM films.

\subsection{Phase and Microstructural Analysis}
X-ray diffraction (XRD) scans were conducted in symmetrical Bragg-Brentano geometry before and after ignition to identify the phases present in the as-deposited multilayers and to analyze phase transformations. A Panalytical X’Pert PRO MPD system was used, employing parallel beam geometry and Cu $K_{\alpha}$ radiation.
To analyze the grain structure of both as-deposited and ignited samples, TEM lamellae were prepared from selected specimens using a dual-beam Helios 5UX system (Thermo Fisher Scientific). Scanning transmission electron microscopy (STEM) and transmission electron microscopy (TEM) were performed using a Talos F200X instrument (Thermo Fisher Scientific) operated at 200 kV and equipped with a SuperX EDX system consisting of four Silicon Drift Detectors (SDD) for elemental content analyses. Additionally, selected area electron diffraction (SAED) was conducted for phase identification.

\subsection{Ignition and Reaction Analysis}
Al/Ni-Ru multilayers, deposited on the Si wafer coated with photoresist, were ignited using a custom-built ignition setup. An Agilent Technologies N6700B Low-Profile MPS mainframe served as the power source, delivering a current pulse through micromanipulators equipped with tungsten probes. The reaction was initiated by applying a voltage of \qty{2}{\volt} and a current of \qty{10}{\milli\ampere}.

The propagation speed of the reaction front and the peak temperature during the self-sustained heat wave were monitored using a high-speed infrared camera (IRCAM Millenium 327k S/M) positioned above the multilayers. Front propagation was recorded at a maximum frame rate of \qty{2.8}{\kilo\hertz}, and the temperature was determined based on the emissivity of the topmost material. The emissivity  for each Ni-Ru composition was calculated using the wavelength-dependent emissivity of pure Ruthenium and Nickel over the range 1.03–4.96\,{\textmu}m. For each wavelength, the emissivity of the multilayer was obtained as a linear combination of the constituent metals, weighted by their atomic fractions in the topmost layer:

\begin{equation}
\varepsilon_\text{mix}(\lambda) = f_\text{Ru} \cdot \varepsilon_\text{Ru}(\lambda) + f_\text{Ni} \cdot \varepsilon_\text{Ni}(\lambda)
\end{equation}
The resulting values were then averaged over the wavelength range to yield a single effective emissivity for each composition, which was used to convert the infrared signal into the actual surface temperature.

\subsection{Molecular Dynamics study of Al/Ni-Ru}
The Molecular Dynamics (MD) simulations were conducted using the Large-scale Atomic/Molecular Massively Parallel Simulator (LAMMPS) package~\cite{PLIMPTON19951}. The Embedded-Atom Method (EAM) potential by Yue et al.~\cite{YUE2020539} was employed to model the interatomic forces between Ni, Al, and Ru atoms. However, the original EAM potential wrongly predicts attraction between Ni and Ru at very short-range separations.

To resolve this issue, the EAM potential was modified using a hybrid potential approach, see Fig.~\ref{fig:mdmlinitial}(d). The Ziegler-Biersack-Littmark (ZBL) potential was overlaid onto the EAM potential to account for short-range repulsive interactions between the Ni and Ru atoms. The hybrid potential reproduces the EAM behavior near the equilibrium separation ($\sim$ 1.9\,{\AA}), maintaining the correct bonding energy and geometry. The ZBL contribution is overlaid beginning at the crossover distance of 1.0\,{\AA} and smoothly switched off by 2.0~{\AA}, ensuring a continuous transition between the repulsive and bonding regimes. At shorter interatomic distances ($r \lesssim$ 1.5\,{\AA}), the potential rises steeply, reflecting the short-range repulsion and correcting the short distance non-physical attraction, seen in the original EAM potential. This smooth crossover ensures continuity of forces, improving numerical stability in molecular dynamics simulations under high-energy conditions such as collisions or displacement cascades. In addition, cohesive energies of Ni--Ru in various crystal structures (L1$_2$, D0$_{22}$, D0$_{19}$, and B2) were calculated, confirming that the hybrid potential preserves the relative structural stability and equilibrium geometries.
The system was designed with fixed dimensions of $L \times d \times h$, where $L$ = 430\,nm represents the length along the $x$-direction, $d$ = 2.8\,nm corresponds to the depth in the $y$-direction, and $h$ = 10\,nm denotes the height in the $z$-direction, equivalent to the bilayer thickness (see FIG.~\ref{fig:mdmlinitial}(a)). Fixed, non-periodic boundaries were applied along the 
$x$-direction, with 86\,nm of vacuum on both sides, preventing atoms from leaving the simulation region. In contrast, periodic boundary conditions were maintained in the $y$- and 
$z$-directions, effectively modeling an infinitely deep and infinitely high multilayer. Such a configuration is comparable to the central region of experimental multilayers, which typically extend over several millimeters in width and consist of at least 50 bilayers.

Al/Ni-Ru multilayers were designed with two compositions: (1) alternating layers of pure Al (50 at.\%) and pure Ni (50 at.\%), and (2) alternating layers of Al (50 at.\%) and a Ni-Ru alloy (Ni 99 at.\%, Ru 1 at.\%) making up the remaining 50 at.\%. The Ru content in the Ni layer was intentionally kept low due to limitations of the interatomic potential. To validate this choice, we computed the enthalpy of mixing for the Ni-Ru system and compared it with literature values. Our analysis showed that when the Ru concentration in the Ni layer exceeds ~4\%, Ru atoms begin to segregate, leading to phase separation that makes the solid solution energetically unfavorable, and the enthalpy of mixing deviates from the expected literature values as quoted by Su \textit{et al.}~\cite{SU2005217}(see FIG.~\ref{fig:mdmlinitial}(c)).  

After the system was created, it was equilibrated in an NPT ensemble with zero pressure and constant temperature $T_0$ = 300\,K for 0.4\,ns. The self-propagating reaction was then initiated by heating the region at $x <$ 50\,nm to 1400\,K, still within the NPT ensemble. Simultaneously, the region at $x >$ 100\,nm was maintained at $T_0$ = 300\,K in an NPT ensemble, while the intermediate region (50\,nm $< x <$ 100\,nm) evolved in an NPH ensemble. The pressure of the barostat was consistently set to 0\,Pa. This setup enabled the creation of a realistic temperature distribution after ignition. The self-propagating reaction was then observed by allowing the system to evolve in an NPH ensemble for 12\,ns.

The front propagation velocity, $v_f$, was determined by recording the position of the reaction front at intervals of 0.2\,ns. The reaction front was defined as the point where the temperature reached 803\,K, corresponding to the melting temperature of Al of the used potential. The temperature profile was obtained using a binning in $x$ using 100 bins. The temperature at the center of each bin was computed as the average temperature within that bin, while linear interpolation was applied between bins. Following previous studies~\cite{fabian2022,fabian2021,fabian2022_2}, a time step of 1 fs was used. Visualization and structural analysis of the system, including atom type identification and crystal structure evaluation, were conducted using the scientific visualization software OVITO~\cite{Stukowski_2010}.

\begin{figure*}
    \centering
    \includegraphics[clip,width=18cm]{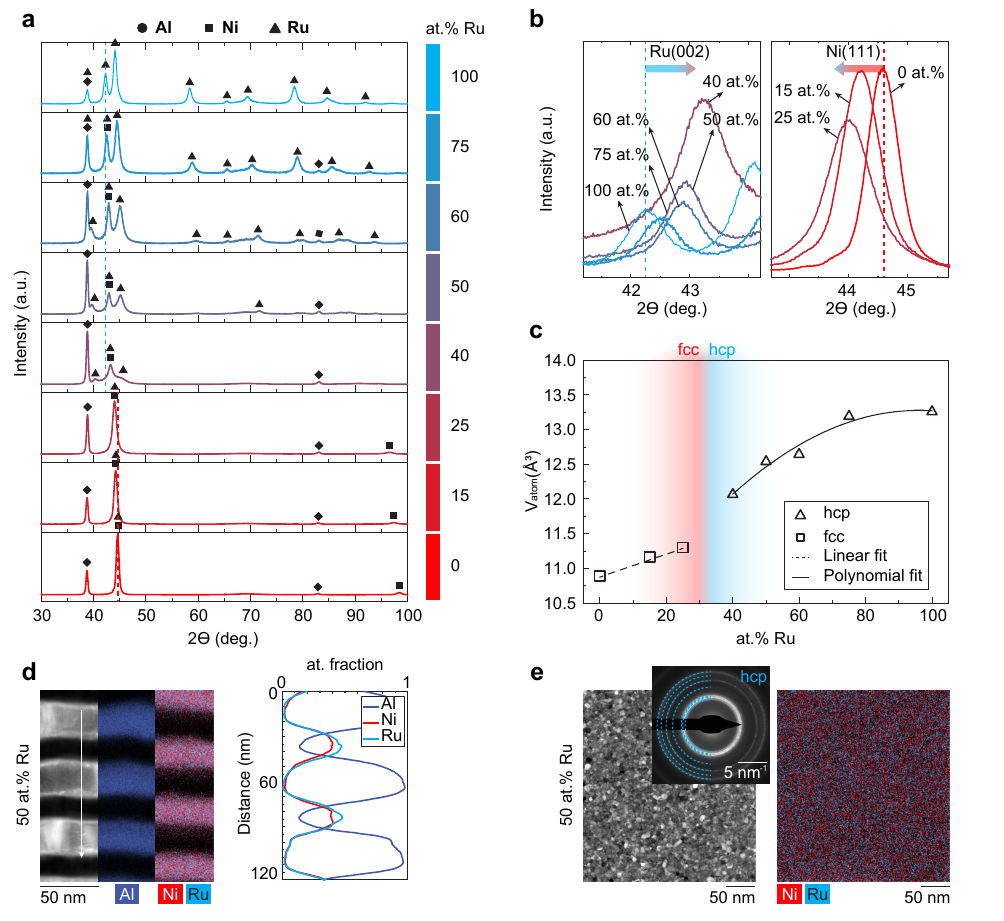}
    \caption{Structural, phase, and chemical fingerprints of as deposited Al/Ni-Ru multilayers as a function of Ru fraction in Ni-Ru layers. (a) XRD patterns of the as-deposited samples from Al/Ni to Al/Ru; the color code denotes the nominal Ru content in the Ni-Ru layer (b) Enlarged region showing shifts of Ru(002) peaks for samples with $\ge 40$ at.\% Ru and Ni(111) peaks for $\le 25$ at.\% Ru, indicating lattice distortion from alloying.  (c) Atomic volume extracted from the fcc and hcp lattice parameters as a function of Ru content. (d) Cross-sectional STEM-EDX maps and corresponding line profile for the 50 at.\% Ru multilayer, confirming homogeneous element distribution and absence of segregation. The shaded band indicates the fcc-to-hcp transition range identified from XRD for the Ni–Ru layer in the as-deposited multilayers (Al remains fcc across the full composition range). (e) Dark-field STEM plan-view of a 20\,nm 50Ni–50Ru layer deposited directly on a TEM grid; the SAED inset exhibits hcp-type diffraction rings, and the EDX overlay indicates nanoscale mixing across the film.}
    \label{fig:struct_comp}
\end{figure*}
    
The maximum temperature behind the propagating reaction front was determined from molecular dynamics simulation data as follows. The temperature profile along the sample, $T(x)$, was recorded at discrete time steps. First, the position of the reaction front was defined as the location where the local temperature reaches a threshold of 803~K. The first time step at which the front crossed $x = 100$\,nm was identified. Then, for all subsequent time steps after this crossing, only the portion of the temperature profile beyond $x = 100$\,nm was considered. The highest temperature within this region, together with its corresponding timestep and position, was extracted as the maximum temperature behind the front. This approach ensures that the reported peak temperature reflects only the portion of the multilayer that has been reached by the propagating reaction front, excluding contributions from regions that were initially heated by the ignition or not yet affected by the front.

\section{Results}
The structural evolution of the as-deposited Al/Ni-Ru multilayers as a function of Ru content in the Ni-Ru layer is shown in FIG.~\ref{fig:struct_comp}. Across the entire composition range, the XRD patterns in FIG.~\ref{fig:struct_comp}(a) show a distinct Al(111) reflection, indicating that the Al layers remain crystalline and unaffected by Ru incorporation within the adjacent Ni-Ru layers. For Ru fractions up to 25 at.\%, the diffraction peaks correspond predominantly to fcc-Ni and shift progressively to lower 2$\Theta$ values with increasing Ru content, indicating incorporation of Ru into the Ni lattice. For higher Ru fractions ($> 25$ at.\%), the main diffraction peaks are associated with hcp-Ru and shift to lower 2$\Theta$ values with increasing Ru content, suggesting a reciprocal incorporation of Ni into the Ru lattice. This trend is highlighted in the magnified view of the main Ru(002) and Ni(111) peaks shown in FIG.~\ref{fig:struct_comp}(b), which clearly illustrates the systematic peak shifts with composition. The corresponding atomic volumes extracted from these reflections are plotted in FIG.~\ref{fig:struct_comp}(c). In the fcc region, the atomic volume, determined from the Ni(111) peak, shows a nearly linear increase with Ru content, consistent with Vegard-like behavior in a substitutional fcc solid solution\cite{jacob2007vegard}. In the hcp region, the atomic volume, calculated using both the Ru(100) and Ru(002) reflections, follows a non-linear, second-order trend. This curvature likely stems from the anisotropic response of the hcp lattice, where solute incorporation affects the \textit{a} and \textit{c} parameters differently, producing a non-ideal evolution of the average atomic volume. Such deviations from ideal Vegard-like behavior are commonly observed in binary solid solutions exhibiting substantial size/lattice mismatch (e.g.,$(>5\%)$)\cite{jacob2007vegard,Danzi2019}. Taken together, these findings show that both the fcc-Ni and hcp-Ru-type layers accommodate substitutional alloying, but the volumetric response is governed by the intrinsic symmetry and elastic anisotropy of their respective crystal structures.

\begin{figure*}
    \centering
    \includegraphics{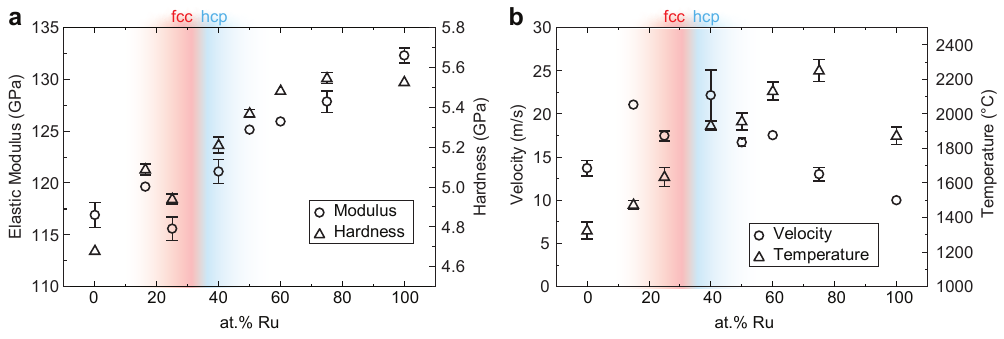}
    \caption{Composition trends in as-deposited mechanical properties and ignition behavior of Al/Ni–Ru multilayers. (a) Elastic modulus and hardness of the as-deposited multilayers as a function of Ru content. (b) Reaction front velocity and reaction temperature as functions of Ru content measured during ignition. The shaded band indicates the fcc-to-hcp transition range identified from XRD for the Ni–Ru layer in the as-deposited multilayers (Al remains fcc across the full composition range). Error bars are omitted where they are smaller than the plotting symbols.}
    \label{fig:mechanicalproperties}
\end{figure*}

STEM-EDX mapping of the 50Ru-50Ni multilayer (FIG.~\ref{fig:struct_comp}(d)) reveals well-defined alternating layers of Al and the co-sputtered Ni–Ru layer. The line scan across the multilayer shows sharp and relatively flat interfaces, indicating that the layer integrity is preserved and that the Ni–Ru layer is compositionally uniform. Dark-field STEM imaging of a 20\,nm thick 50Ru–50Ni layer deposited on a TEM grid (FIG.~\ref{fig:mechanicalproperties}(e)) shows a nanocrystalline microstructure, while the corresponding SAED pattern exhibits broad diffraction rings consistent with crystallites of about 10\,nm in size and lattice strain due to Ni–Ru alloying. STEM-EDX mapping confirms a homogeneous distribution of Ni and Ru within the layer, supporting the formation of a metastable solid solution under non-equilibrium sputtering conditions.

The elastic modulus and hardness of the as-deposited multilayers as a function of Ru content are shown in FIG.~\ref{fig:mechanicalproperties}(a). Both quantities increase overall with higher Ru concentration, rising from 116.9 $\pm$ 1.2\,GPa to 132.3 $\pm$ 0.7\,GPa for the modulus and from 4.68 $\pm$ 0.01\,GPa to 5.53 $\pm$ 0.01\,GPa for the hardness. A small drop appears at 25 at.\% Ru, after which both values continue to rise steadily. This deviation near the fcc-hcp transition likely reflects a change in crystal symmetry and local bonding environment within the Ni-Ru layers, consistent with the structural shift observed in the XRD results. The progressive stiffening and hardening with Ru addition are consistent with solid-solution strengthening in the Ni-rich fcc Ni–Ru layers at low Ru contents, together with an increasing contribution from the Ru-rich hcp structure at higher Ru contents, reducing the elastic deformability of the lattice\cite{yang2022strengthening}.

The reaction velocities and peak temperature for the ignited multilayers are plotted in FIG.~\ref{fig:mechanicalproperties}(b). In the Ru-free multilayers the reaction propagates at $13.7 \pm 0.9\ \mathrm{m/s}$ with a peak temperature of about 1300°C. Introducing Ru initially accelerates the reaction, reaching $21 \pm 0.5\ \mathrm{m/s}$ at 15 at.\%Ru, while the peak temperature increases to about 1470°C. With further Ru addition the velocity becomes non-monotonic: it decreases at 25 at.\% Ru ($17.5 \pm 0.6\ \mathrm{m/s}$) and then reaches its maximum at 40 at.\% Ru ($22.2 \pm 3\ \mathrm{m/s}$), close to the fcc-hcp transition region. Beyond this composition the propagation progressively slows down, dropping to $13 \pm 0.8\ \mathrm{m/s}$ at 75 at.\% Ru and $10 \pm 0.3\ \mathrm{m/s}$ at 100 at.\% Ru. In contrast, the peak temperature continues to rise across the transition and within the hcp regime, increasing from about 1900°C at 40 at.\% Ru to a maximum of 2250°C at 75 at.\% Ru, before decreasing again to about 1870°C at 100 at.\% Ru. The Ru-rich limit (100 at.\% Ru) yields both front velocity and peak temperature that are comparable to those reported by Woll \textit{et al.}\cite{Woll2016} for binary Al/Ru reactive multilayer. Overall, the highest velocity coincides with the structural crossover, whereas the temperature maximum is shifted to higher Ru contents, indicating a decoupling between propagation kinetics and the peak thermal output at elevated Ru levels. 

\begin{figure*}
    \centering
    \includegraphics{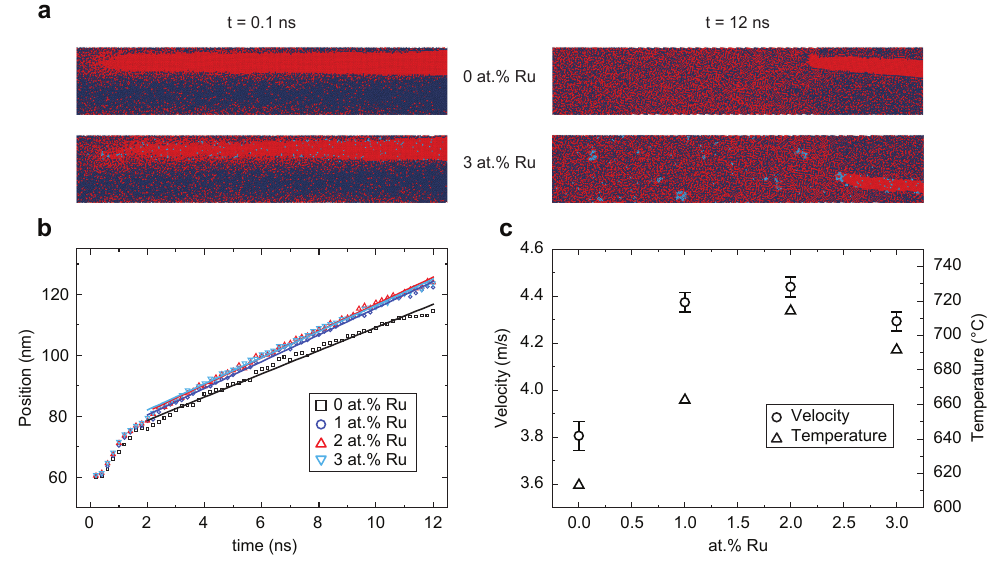}
    \caption{Composition trends in reaction-front morphology and propagation kinetics from molecular-dynamics simulations of Al/Ni multilayers containing dilute Ru additions. (a) Representative MD snapshots comparing the binary Al/Ni stack (top) with the Ru-alloyed system at 3 at.\% Ru (bottom) at t=0.1\,ns t=12\,ns (Al: blue; Ni: red; Ru: cyan). The binary Al/Ni system exhibits a continuous reaction front, while the addition of Ru results in dispersed nucleation events and a more disordered reacted region. (b) Reaction-front position as a function of time for 0–3 at.\% Ru. (c) Extracted front velocity together with the corresponding characteristic reaction temperature from the simulations plotted versus Ru concentration.}
    \label{fig:MD_results}
\end{figure*}

To further clarify these experimental trends, MD simulations were performed for Al/Ni and Al/Ni-Ru multilayers containing up to 3 at.\% Ru in the Ni layer. FIG.~\ref{fig:MD_results}(a) presents the reaction progression over time for the binary Al-Ni system and the Ru-alloyed Al-Ni system containing 3 at.\% Ru in the Ni layer. At $t=$0.1\,ns, both systems exhibit a sharp compositional interface between the Al (blue) and Ni (red) layers, with Ru atoms (green) initially confined to the Ni-rich region. As the reaction progresses, interdiffusion leads to the formation of a body-centered cubic (BCC) phase in both compositions. Structural analysis confirms that the final simulated configuration corresponds to the ordered B2-AlNi phase, which is the most stable phase for the interatomic potential used in this work, with a formation energy of $-4.53$\,eV/atom. The Ru concentration used here ($\leq 3$ at.\%) is below the segregation threshold of the used potential; for higher Ru concentration the potential wrongly predicts Ru segregation.

The MD-simulated time evolution of the reaction front position $x(t)$ is shown in Fig.~\ref{fig:MD_results}(b). The slope of each curve corresponds to the propagation velocity, which is obtained by linear fitting. The resulting front velocities are plotted in Fig.~\ref{fig:MD_results}(c) as a function of Ru content. For the binary Al-Ni multilayer (0~at.\% Ru), the velocity is 
$3.81 \pm 0.06~\mathrm{m\,s^{-1}}$. The addition of Ru leads to a systematic increase in propagation rate, reaching 
$4.37 \pm 0.04~\mathrm{m\,s^{-1}}$ at 1~at.\% Ru and a maximum value of 
$4.44 \pm 0.04~\mathrm{m\,s^{-1}}$ at 2~at.\% Ru, corresponding to an enhancement of approximately 17\% relative to the Ru-free system. At 3~at.\% Ru the velocity decreases slightly to 
$4.29 \pm 0.04~\mathrm{m\,s^{-1}}$, but still remains higher than the binary case. This demonstrates that Ru accelerates the intermixing process. 
The small decrease at 3~at.\% Ru is not considered significant, as the measured velocity remains within the uncertainty range of the 2~at.\% value.

In the MD simulations, the maximum temperature behind the propagating reaction front also increases with Ru content, reflecting more energetic transformations and enhanced intermixing (see FIG.~\ref{fig:MD_results}(c)). The peak temperature rises from approximately 887~K for the Ru-free system to 936~K and 988~K for 1 at.\% and 2 at.\% Ru, respectively, and slightly decreases to 965~K at 3 at.\% Ru, consistent with the small reduction in front velocity.

\begin{figure*}
    \centering
    \includegraphics{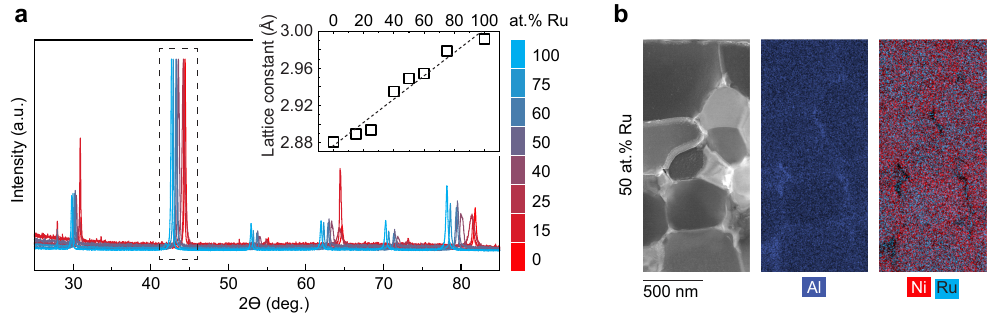}
    \caption{Composition trends in phase formation after ignition and in the resulting microstructure of Al/Ni–Ru multilayers.(a) XRD patterns of the multilayer samples after ignition for the full Ru series, showing reflections of a B2 (NiAl-type) product for all initial Ni–Ru layer compositions; the boxed region highlights the strongest reflection, which shifts systematically with Ru content. Inset: lattice parameter of the B2 product derived from the principal reflections, plotted versus nominal Ru fraction in the starting Ni–Ru layers (b) Representative post-ignition microstructure for the 50 at.\% Ru sample together with STEM/EDX elemental maps, where the Al signal and the combined Ni+Ru distribution visualize the spatial partitioning of the reacted material at the sub-micrometer scale.}
    \label{fig:xrdignitedsystems}
\end{figure*}

FIG.~\ref{fig:xrdignitedsystems}(a) shows the experimental XRD patterns of the multilayers after ignition. Independently of the initial Ru content in the as-deposited state, all samples exhibit reflections corresponding to the ordered B2 NiAl-type phase, indicating the formation of a single intermetallic phase upon reaction. The most intense peak appears near 43°, and its position gradually shifts to lower diffraction angles with increasing Ru content. This continuous peak displacement points to a systematic lattice expansion associated with Ru incorporation. This is quantified in the inset: the lattice parameter increases nearly linearly from $2.88\,\text{\AA}$ at 0 at.\% Ru to $2.99\,\text{\AA}$ at 100 at.\% Ru, consistent with Vegard-like behavior expected for substitutional incorporation of the larger Ru atoms on the Ni sublattice of B2 NiAl. In the reacted 50 at.\% Ru sample (Fig.~\ref{fig:xrdignitedsystems}(b)), the STEM image reveals a coarse polycrystalline microstructure with grains on the order of a few hundred nanometers. The Al EDX map shows an intergranular Al-enriched network that outlines many grains, suggesting that Al is retained along grain boundaries and triple junctions after ignition. Conversely, the Ni–Ru overlay indicates that Ni and Ru are concentrated primarily within the grains, consistent with the grains corresponding to the B2 AlNiRu product identified by XRD. Notably, the Ni/Ru signal within individual grains is not homogeneous; instead, it exhibits local fluctuations in relative Ni vs.\ Ru intensity, indicative of residual compositional heterogeneity on the mapped length scale (and/or thickness effects). Overall, the EDX maps therefore suggest that, while XRD is dominated by the B2 phase, the reacted microstructure retains an Al-enriched intergranular component and intragranular Ni/Ru variations, pointing to incomplete chemical homogenization under the rapid, non-equilibrium ignition conditions.

\section{Discussion}

The incorporation of Ru into the Ni layers of the Al/Ni multilayers affects the structural, mechanical, and reactive behavior of the system. XRD and TEM/STEM-EDX analyses reveal that Ru substitutes for Ni in the lattice, producing a lattice expansion and peak shift consistent with solid-solution formation. The persistence of crystalline Al reflections across all compositions indicates that Al layers remain structurally stable, while Ni-Ru layers accommodate most of the lattice strain.

The structural change from an Ni-like fcc state toward a Ru-like hcp state is bounded in our dataset by the 25 at.\% and 40 at.\% Ru samples. While we do not have intermediate XRD runs, several observations support that the fcc-to-hcp crossover occurs in this interval. The 25 at.\% sample retains predominantly fcc-type reflections and shows a small mechanical dip, whereas the 40 at.\% sample exhibits features more consistent with hcp-rich character and recovered hardening. Together with the continuous evolution of lattice parameters and the known Ni–Ru phase behavior reported in literature, these end-member data imply that mixed stacking (faulted/fcc + hcp regions) is likely to develop between 25 and 40 at.\% Ru. We therefore phrase the transition as an inferred compositional window rather than a directly measured two-phase region: this wording is borne out by the concordance of diffraction, mechanical and microstructural indicators and by previous reports that Ru lowers stacking-fault energy and promotes local hcp stacking in Ni-Ru alloys~\cite{van1990structural,Cheng02112023,Kitashima2015,breidi2024first}. Moreover, far-from-equilibrium ion mixing of Ni/Ru has been shown to generate a Ni-enriched hcp phase and a Ru-enriched fcc phase\cite{li2002nonequilibrium}.

This structural interpretation is consistent with the microscopy one: the 50Ni-50Ru layer remain compositionally homogeneous at the STEM-EDX lenght scale (FIG.~\ref{fig:struct_comp}(e)) and and preserve sharp AL/(Ni-Ru) interfaces (FIG.~\ref{fig:struct_comp}(d)), but the 50Ni-50Ru layer is nanocrystalline with broad SAED rings, consistent with high defect density and significant microstrain from substitutional mismatch. Such a strained nanocrystalline state is typical of sputter-deposited, non-equilibrium solid solutions\cite{chen2025nanoscale} and is particularly important here because it preconditions the reactive stack with a high density of short-circuit diffusion paths (grain boundaries, faulted planes) and stored elastic energy.  

This inferred phase transition region provides a plausible explanation for the non-monotonic trend observed in mechanical properties. The increase in hardness and modulus with Ru addition up to 25 at.\% reflects classical solid-solution strengthening, where substitutional Ru atoms distort the Ni lattice and hinder dislocation motion\cite{neumeier2011influence,fleischer1963substitutional}. The dip in both properties within the 25–40 at.\% window coincides with the onset of structural instability and stacking disorder, which reduce the effective load-bearing capability of the multilayer. Beyond 40 at.\% Ru, the recovery of hardness and elastic modulus is consistent with the Ni–Ru layers becoming Ru-rich and adopting an hcp-dominated structure with intrinsically high stiffness.\cite{wang2012ideal}. In hcp lattices, plastic accommodation is more strongly constrained because deformation along the c-axis relies on ⟨c+a⟩ slip, which requires high critical resolved shear stress and exhibits limited self-multiplication, thereby increasing resistance to plastic flow and contributing to higher hardness\cite{zou2025playing}.

Reaction-front propagation in Al/transition-metal multilayers is set by the competition between diffusion-controlled mixing in a narrow interfacial zone and the rate at which the released heat can sustain that zone as it advances, so front speed is exceptionally sensitive to the effective mixing length and any deposition-induced premixing. Time-resolved synchrotron XRD during rapid heating of Al/Ni foils shows that Ni begins to diffuse into the Al layer already at ~271 °C and that atomic diffusion governs ignition under these conditions\cite{neuhauser2020analysis}, analogously, Ru/Al reactions proceed by Ru dissolving into Al\cite{pauly2015role}. Within this framework, the initial acceleration at low Ru is consistent with defect-assisted enhancement of Ni/Ru transport into Al, while the non-monotonic response near the inferred fcc-hcp crossover plausibly reflects a maximum in faulted/nanocrystalline short-circuit pathways that transiently lower the kinetic barrier for mixing. At higher Ru, the system trends toward the Ru/Al limit, where thermochemical analysis indicates a substantially higher adiabatic temperature than Ni/Al and experiments report about 2000~°C peak temperatures yet only about $10  \mathrm{m/s}$ front velocities: an explicit reminder that increased thermal output does not guarantee faster fronts once intermixing becomes rate-limiting—consistent with the observed decoupling between a velocity maximum near the crossover and a temperature maximum deeper in the Ru-rich regime\cite{Woll2016}.

Atomistic simulations reinforce this picture at the low-Ru limit: adding $\leq$3 at.\% Ru in the Ni layer increases reaction-front velocity and peak temperature. This behavior can be explained by the combined effects of modification of the local chemical driving force and increase of the diffusivity. This interpretation is directly supported by diffusion-couple MD annealing (SFig.~1, 700\,K, 10\,ns), where Ni penetration into the Al side is markedly enhanced in the presence of 3 at.\% Ru, while Al transport into the Ni-rich region remains comparatively limited. The asymmetric interdiffusion is therefore preserved, but the effective mixing length on the Al side increases with Ru addition. 

The small drop in speed and peak temperature observed at 3 at.\% Ru is not physically significant, as it falls within the uncertainty of the 2 at.\% value. Higher Ru concentrations were not investigated, since the EAM potential becomes unreliable above ~4 at.\% Ru, leading to artificial segregation and deviations in mixing enthalpy (Fig.~\ref{fig:mdmlinitial}(c)). Thus, the physically meaningful simulated trend is the continuous increase in velocity up to 2 at.\% Ru. The experimental observation that modest Ru additions accelerate fronts, together with the simulated transition toward multi-site reactivity, echoes broader mechanistic insights from reactive multilayer literature: early-stage solid-state reactions at interfaces and internal boundaries can dominate ignition and early propagation, and chemistry that increases the density or reactivity of such boundaries reduces the temperature/enthalpy “overhead” needed to sustain a traveling front.

Despite the strong kinetic and thermal changes, post-ignition XRD shows a single dominant ordered B2 phase for all starting Ru contents with a nearly linear lattice-parameter increase, indicating substitutional incorporation of Ru on the Ni sublattice of NiAl-type B2. This outcome is thermodynamically plausible because NiAl and RuAl share the same B2 structure and are predicted/assessed to form extensive B2 solid solutions at high temperature without a miscibility gap above 1273~K, so rapid reactions can converge on a single B2 field even under non-equilibrium conditions\cite{hallstrom2008thermodynamic,Woll2016}. At the same time, the reacted microstructure (coarse grains with an Al-enriched intergranular network and intragranular Ni/Ru fluctuations) indicates that the reaction freezes in chemical heterogeneity. The present EDX signatures are therefore best interpreted not as evidence against single-phase B2 formation, but as a kinetic imprint of the reaction’s brevity: XRD is dominated by the crystallographically coherent B2 grains, while fast solidification and limited diffusion trap Al-rich intergranular regions and local Ni/Ru partitioning within grains. 
Supplementary STEM/EDX (SFig.~2) shows that this quenched chemical contrast is not monotonic with composition. The Ru end-member is comparatively uniform at the sub-micrometer scale, consistent with rapid consumption of Al into a RuAl-type B2 product. At 75 and 25 at.\% Ru, an Al-enriched matrix-like fraction becomes apparent around the coarse B2 grains, whereas at 60 at.\% Ru the continuous matrix is instead enriched in Ni. In the Ru-rich regime (e.g., 75 at.\% Ru), Ni additionally enriched along parts of the grain boundaries. These trends point to a composition-dependent partitioning of the last-reacted/last-solidified fraction during the reaction, which is retained upon rapid quenching.
Furthermore, it has been shown by first-principle calculations that Ru preferentially occupies the Ni sublattice in B2-NiAl, while Ni favors the Ru sublattice in B2-RuAl\cite{prins2007defect}. This sublattice selectivity provides a microscopic basis for a Vegard-like lattice expansion while preserving B2 symmetry.

\section{Conclusion}

Taken together, our results establish Ru alloying of the Ni sublayer as a robust and continuous design lever for Al/Ni reactive multilayers, enabling controlled tuning from Ni-rich to Ru-rich sublayers while preserving sharp multilayer architectures, homogeneous elemental distributions, and nanocrystalline domains with a predictable lattice-parameter evolution. The as-deposited structural trends indicate an inferred fcc to hcp crossover between 25 and 40 at.\% Ru, where stacking disorder and microstrain likely peak, providing a plausible origin for the observed non-monotonic hardness and modulus response. Mechanistically, the MD simulations indicate that Ru primarily reprograms the reaction pathway, promoting multi-site nucleation and more distributed intermixing, so that modest changes in apparent front velocity should be interpreted in terms of altered front morphology and heat-release localization rather than a simple kinetic slowdown. Experimentally, reaction-front measurements show a non-monotonic dependence of propagation on Ru content, front speeds pass through a maximum while peak temperatures continue to rise toward the Ru-rich regime, highlighting that increased thermal output does not necessarily translate into faster fronts once intermixing becomes rate-limiting. More broadly, because Ni/Al reactive multilayers are widely used as ultrafast, localized heat sources (e.g., for joining), the ability to tune reaction-front characteristics without changing the ultimate B2 product provides a practical route to engineer ignition robustness and thermal delivery for targeted applications. Finally, extending the simulations to experimentally matched Ru concentrations will require improved interatomic descriptions, but the combined experimental–computational evidence already demonstrates composition-enabled control over both the parent multilayer state and the reactive transformation pathway within a single, scalable material platform.

\begin{acknowledgments}
Electron microscopy analysis was performed at ScopeM, the microscopy platform of ETH Zürich. Sputter deposition of the multilayer systems was performed at FIRST, the clean room facilities of ETH Zürich. Computational resources were provided by the e-INFRA CZ project (ID: 90254), supported by the Ministry of Education, Youth and Sports of the Czech Republic. This work has been supported by the project INTER-COST No. LUC24093 funded by the Ministry of Education, Youth and Sports of the Czech Republic. The authors are grateful to Fabian Schwarz for his constructive feedback on the manuscript.

\end{acknowledgments}

\section*{AUTHOR DECLARATIONS}
\subsection*{Conflict of Interest}
The authors have no conflicts to disclose.

\subsection*{Authors Contributions}
\textbf{N.T.}: Conceptualization (lead); Data curation (lead); Formal analysis (lead); Investigation (lead); Methodology (lead); Validation (lead); Visualization (lead); Writing – original draft (lead); Writing – review \& editing (lead). \textbf{A.Y.}: Conceptualization (supporting); Data curation (lead); Formal analysis (lead); Investigation (lead); Methodology (lead); Validation (lead); Visualization (lead); Writing – original draft (lead); Writing – review \& editing (lead). \textbf{J.F.}: Conceptualization (supporting);  Project administration (lead); Resources (lead); Supervision (lead); Writing–review and editing (supporting). \textbf{R.S.}: Conceptualization (supporting);  Project administration (lead); Resources (lead); Supervision (lead); Writing–review and editing (supporting).

\section*{Data Availability Statement}
The data that support the findings of this study are available from the corresponding author upon reasonable request.

\section*{references}

\end{document}